\documentclass[a4paper,twocolumn, nofootinbib,superscriptaddress,aps,prm, 14pt,eqsecnum,notitlepage,showkeys]{revtex4-1}
\usepackage{multirow}

\usepackage{bm}
\usepackage{dcolumn}
\usepackage{hyperref}
\usepackage{gensymb}
\usepackage{amsmath}
\usepackage{dcolumn}    % Align table columns on decimal point
\usepackage{ifpdf}
\usepackage{amssymb,lineno,amsfonts}
\usepackage{graphicx}   % Include figure files
\usepackage{bm}         % bold math
\usepackage{bbm}
\usepackage{mathrsfs}
\usepackage{upgreek}
\usepackage{mathtools}
\usepackage{epstopdf}
\usepackage{setspace}
\usepackage{hyperref}
\usepackage{natbib}
\usepackage{esvect}
\usepackage{amsmath}
\usepackage[usenames,dvipsnames]{xcolor}
\definecolor{med-blue}{RGB}{25,25,112}
\hypersetup{colorlinks, linkcolor={blue},citecolor={blue}, urlcolor={blue}}
\usepackage{times }
\usepackage [english]{babel}
\usepackage [autostyle, english = american]{csquotes}
\MakeOuterQuote{"}
\begin{document}
\title{Absence of a multiglass state in some transition metal doped quantum paraelectrics}
\author{Charu Garg}
\author{Jitender Kumar}
\affiliation{Department of Physics, Indian Institute of Science Education and Research}
\author{Sunil Nair}
\affiliation{Department of Physics, Indian Institute of Science Education and Research}
\affiliation{Centre for Energy Science, Indian Institute of Science Education and Research,\\ Dr. Homi Bhabha Road, Pune, Maharashtra-411008, India}
\date{\today}
\begin{abstract} 
	We critically investigate the purported existence of a multiglass state in the quantum paraelectrics SrTiO${_3}$ and KTaO${_3}$ doped with magnetic 3d transition metals. We observe that the transition metals have limited solubility in these hosts, and that traces of impurity magnetic oxides persist even in the most well processed specimens. Our dielectric measurements indicate that the polar nano-regions formed as a consequence of doping appear to lack co-operativity, and the associated relaxation process exhibits a thermally activated Arrhenius form. At lower temperatures, the dielectric susceptibility could be fit using the Barrett's formalism, indicating that the quantum-paraelectric nature of the host lattices are unaltered by the doping of magnetic transition metal oxides. All these doped quantum paraelectrics exhibit a crossover from the high temperature Curie-Weiss regime to one dominated by quantum fluctuations, as evidenced by a $T{^2}$ dependence of the temperature dependent dielectric susceptibility. The temperature dependence of the magnetic susceptibility indicate that magnetic signatures observed in some of the specimens could be solely ascribed to the presence of impurity oxides corresponding to the magnetic dopants used. Hence, the doped quantum paraelectrics appear to remain intrinsically paramagnetic down to the lowest measured temperatures, ruling out the presence of a multiglass state.  
	\end{abstract}
	\maketitle

\section{Introduction}

Multiferroics - materials with co-existing magnetic and polar orders - have been at the focus of extensive theoretical and experimental investigations. Initially thought to be phenomena which are inimical to each other, a variety of avenues are now known, by which a system can be tailored to exhibit both these orders  \cite{fiebig}. Not surprisingly, advances in this area of research have been fuelled by the continuous availability of new materials, which exhibit coupling between electric and magnetic order parameters. Multiferroics can be broadly divided in two different classes: Type -I multiferroics, where magnetic and polar orders arise from independent microscopic origins, and Type-II multiferroics, where a non-trivial magnetic order facilitates ferroelectricity \cite{cheong,tokura}. An additional variant has also been reported - that in which both the magnetic and polar orders are frozen and these \emph{multiglasses} are purported to be characterized by the co-existence of a magnetic spin/cluster glass with a polar cluster glass phase \cite{kleemann1,kleemann2}. 
  
The genesis of the area of multiglasses can be traced to the report of the simultaneous occurrence of a non-ergodic polar and magnetic glass state in the doped quantum parelectric Sr${_{0.98}}$Mn${_{0.02}}$TiO${_3}$ \cite{multiglass1, multiglass2}. It was suggested that the Mn${^{2+}}$ ions which replace the Sr${^{2+}}$ species in SrTiO${_3}$ undergo off-centre displacements from their mean positions, creating electric dipoles. Such polar clusters then undergo a low temperature transition to a frozen relaxor-like state at $T{_G}$ = 38K as was evidenced by a power law dependence of the frequency dependent dielectric susceptibility \cite{multiglass1}. Since Mn${^{2+}}$ is also magnetic ($ S=5/2$), they couple to each other via frustrated antiferromagnetic superexchange interactions, and below $T{_G}$ the freezing of polar clusters also triggers the freezing of spin degrees of freedom. This effect is also augmented by the presence of a finite Magneto-electric coupling, which couples the polar and spin degrees of freedom, resulting in a \emph{magnetoelectric multiglass} state in Sr${_{0.98}}$Mn${_{0.02}}$TiO${_3}$. Though the closely related K${_{0.97}}$Mn${_{0.03}}$TaO${_3}$ system was also reported to harbor such a state \cite{kleemann3}, it was later described as a spin cluster glass where the interacting polar clusters fail to condense into a glassy state \cite{cluster}. The coexistence of a magnetic and polar glass like states was also reported in the double perovskite La${_2}$NiMnO${_6}$ \cite{dd1} and the delafossite CuCr${_{0.5}}$V${_{0.5}}$O${_2}$ \cite{delafossite} systems, with the anti-site disorder (between the Ni-Mn and Cr-V species respectively) being responsible for the observed multiglass state. 

\begin{figure*}
\centering
	\includegraphics[scale=0.2]{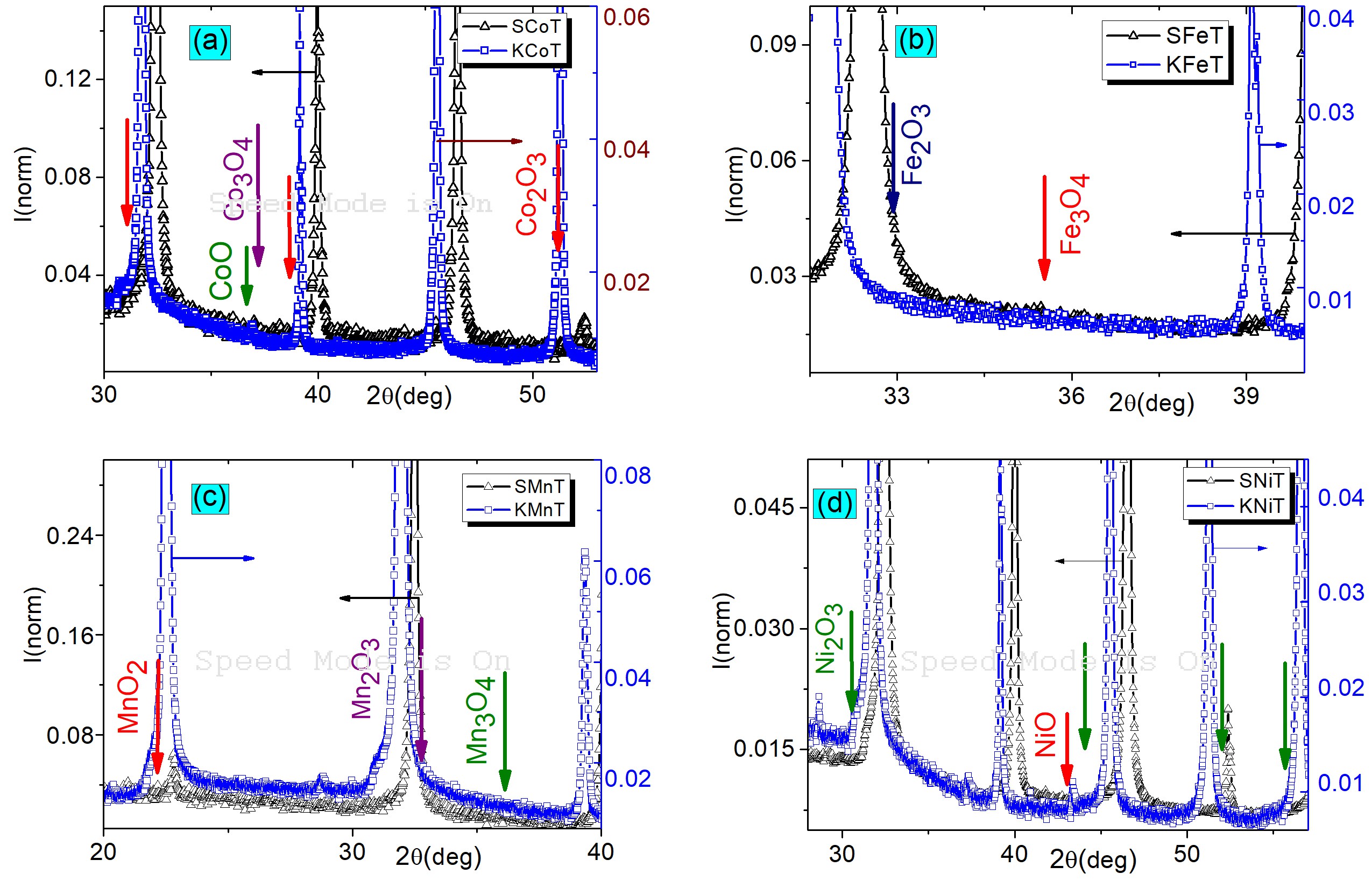}
	\caption{Room temperature x-ray diffraction data of the KTaO${_3}$ and SrTiO${_3}$ specimens doped with  (a) Co, (b) Fe, (c) Mn and (d) Ni. The 2$\theta$ ranges have been selected such that it encompasses the region where the most intense peak of the possible magnetic impurities (if any) can be clearly seen.}
	\label{Fig1} 
\end{figure*}

A number of experimental signatures were used to characterize this multiglass state in Sr${_{0.98}}$Mn${_{0.02}}$TiO${_3}$ \cite{kleemann1,kleemann2}. These included (i) magnetic irreversibility in the zero field cooled and field cooled measuring protocols, (ii) the observation of a frequency dependent ac susceptibility which appears to coincide with the polar freezing temperature, (iii) a power law fit to the peak temperature of the real part of the dielectric permittivity $\epsilon'(T)$, (iv) memory effects in both the polar and magnetic sectors and (v) a peak in the non-linear dielectric susceptibility \cite{multiglass1,multiglass2,kleemann3}. \\
However, there have been a number of subsequent reports which have raised doubts about the validity of the multiglass scenario in this system \cite{choudhury2011tuning}, and in the transition metal doped quantum paraelectrics in general. The significant observations which are not commensurate with the existence of a multiglass state included (i) the absence of magnetic freezing in specimens synthesized using different synthesis procedures like high energy ball milling \cite{afm1}, and the oxalate precipitate method \cite{PREETHIMEHER2012296} (ii) Transmission Electron Micrographs indicating that the magnetic anomaly was only observed in those specimens where a network of structural defects were present \cite{afm1}, The inference therein was that the existence of structural defects was an indicator of the fact that the dopant (Mn) was not homogeneously distributed in the host lattice, and that the segregation of these magnetic dopants (and possibly their oxides) was responsible for the observed magnetic signatures. (iii) Investigations with careful site specific Mn substitutions indicating that Sr${_{0.98}}$Mn${_{0.02}}$TiO${_3}$ does not exhibit any magnetic ordering/anomalies, where as Mn co-doped in both the Sr and Ti sites exhibit glassy magnetic properties of extrinsic origin\cite{choudhury2011tuning}. Doping the sample at Ti site left the system in a paraelectric state whereas doping at Sr ( and Sr and Ti simultaneously) site lead to the formation of relaxor like state. The magnetic ordering  was proven to be decoupled from the dielectric one, thus ruling out the possibility of a multiglass phase in these systems. Moreover, it was explicitly stated that the magnetic anomaly observed in the Sr and Ti site co-doped Mn was due to presence of Mn$_3$O$_4$ impurities in small quantities. (iv) the absence of a frequency dependence in the measured non linear dielectric susceptibility \cite{kleemann3}.  

The main point of contention pertained to freezing in the magnetic sector, and the possibility that the observed glassy magnetic signatures arises from a small amount of the spinel Mn${_3}$O${_4}$, which is reported to exhibit a ferrimagnetic transition at $\approx$ 43K \cite{dd1,afm1}. At the small doping percentages under consideration here, these trace i
mpurities are not easily discernible from routine x-ray diffraction measurements. This problem is accentuated by  the fact that Mn${_3}$O${_4}$ can dissolve small amounts of TiO${_2}$ (which is typically used as a raw material in the synthesis of Sr${_{0.98}}$Mn${_{0.02}}$TiO${_3}$) \cite{afm1}, and could also exhibit finite size effects when it exists in the form of small impurity clusters within the perovskite matrix, thus exhibiting a sample dependent variation of the observed (glass-like) magnetic transition temperature. The pristine Mn${_3}$O${_4}$ phase is also reported to be a magnetodielectric, which exhibits a number of closely spaced spin re-orientation transitions with clear dielectric anomalies corresponding to these different transitions \cite{ram,suzuki}. 

The fact that Mn${_3}$O${_4}$ exhibits a phase transition in close vicinity to the energy scale of the relaxor state in these doped quantum paraelectrics should mean that this ambiguity should be unique to Mn doping alone. Other magnetic transition metal dopants (Co, Fe or Ni) could offer a means of evaluating the existence of a multiglass state in the doped quantum paraelectrics, since the transition temperatures associated with the possible trace impurities (CoO, Co${_3}$O${_4}$, Fe${_2}$O${_3}$, Fe${_3}$O${_4}$, NiO, or Ni${_2}$O${_3}$) are very different from the temperature range of interest here. However, to the best of our knowledge there has been no attempt to evaluate the feasibility of a magnetically frozen state in quantum paraelectrics doped with different magnetic transition metal dopants. Here, we report an investigation of specimens of Sr${_{0.98}}$$M{_{0.02}}$TiO${_3}$ (SMT) and K${_{0.97}}$$M{_{0.03}}$TaO${_3}$ (KMT), with $M$ = Mn, Co, Fe, or Ni, synthesized using the solid state ceramic route. These samples are characterized using x-ray diffraction, energy dispersive x-ray analysis, and temperature dependent dielectric and magnetization measurements. Though all the doped specimens exhibit frequency dependent features in the dielectric loss, our data does not indicate the presence of a frozen polar state in any of these systems. Moreover, all the observed magnetic signatures can be clearly attributed to arise from the presence of oxides of the magnetic dopants used.  Thus, our observations suggests that transition metal doped quantum paraelectrics clearly do not harbor an intrinsic glassy  state in either the polar or the magnetic sectors.  

  \section{Experimental}
Polycrystalline specimens of the Sr${_{0.98}}$$M{_{0.02}}$TiO${_3}$ and K${_{0.97}}$$M{_{0.03}}$TaO${_3}$ series were synthesized by the standard solid state reaction technique. For the SMT series, stoichiometric amounts of the preheated reagents were treated twice at 1150$^{\circ}$C followed by pelletizing and a final treatment at 1500$^{\circ}$C for 24 hours.. For the KMT series, the protocol was slightly different, and stoichiometric amounts of the preheated reagents were mixed and ground thoroughly in a glove box for 3 hours, followed by a heat treatment at 1000$^{\circ}$C. This mixture was then reground and pelletized before being subjected to a final heat treatment at 1000$^{\circ}$C for 24 hours. Multiple batches of samples were synthesized, and only the best ones were used for further investigations. 

\begin{figure}
  \includegraphics[width=\linewidth]{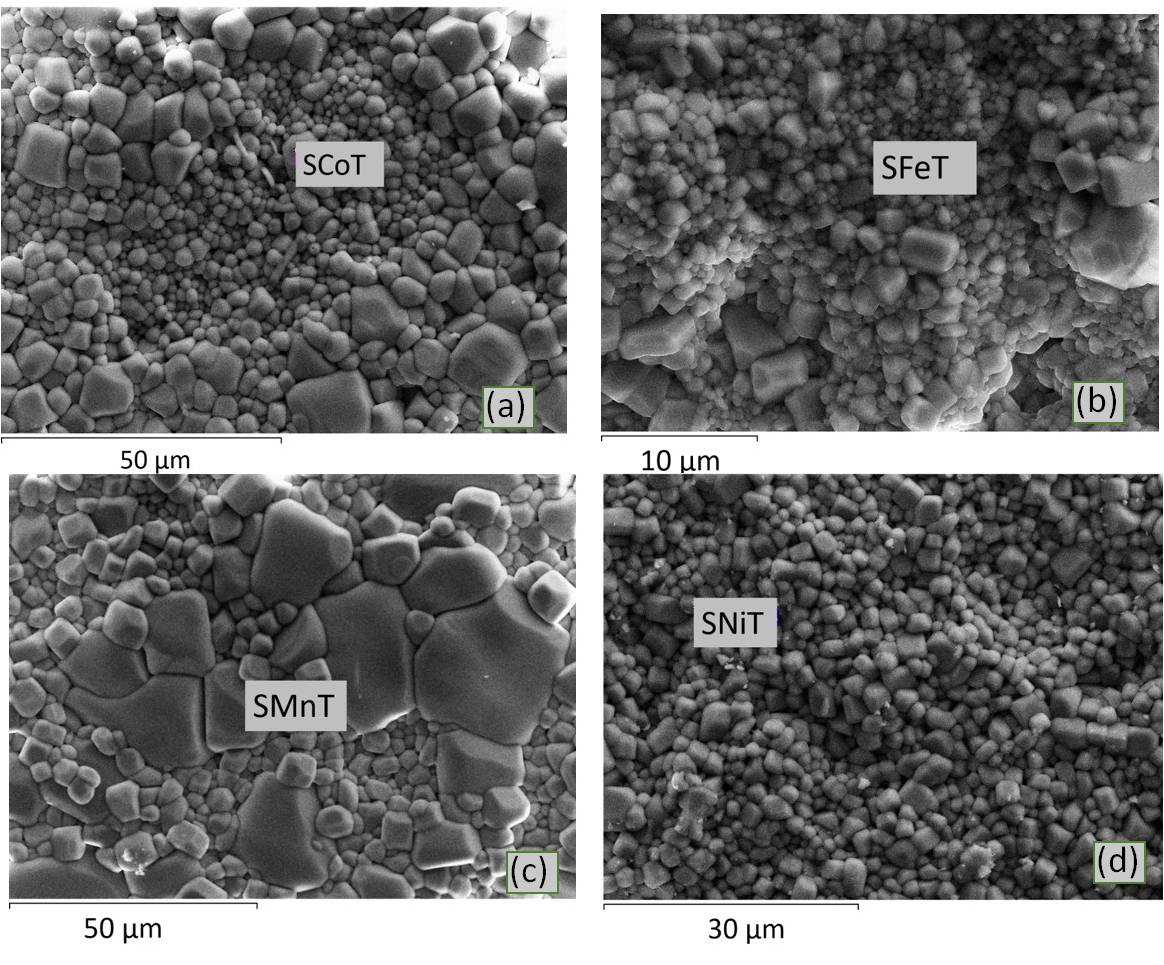}
  \caption{Scanning electron micrographs of SrTiO${_3}$ specimens doped with (a)Co, (b)Fe, (c)Mn and (d)Ni. Note that all the scans were performed on the piece of a pellet which gets manifested in the form of high density and homogeneity of grains. Scans were performed on various regions of the pellet and the obtained elemental contribution was consistent with the expected ratio.}
  \label{Fig2}
\end{figure}

\begin{figure}
  \includegraphics[width=\linewidth]{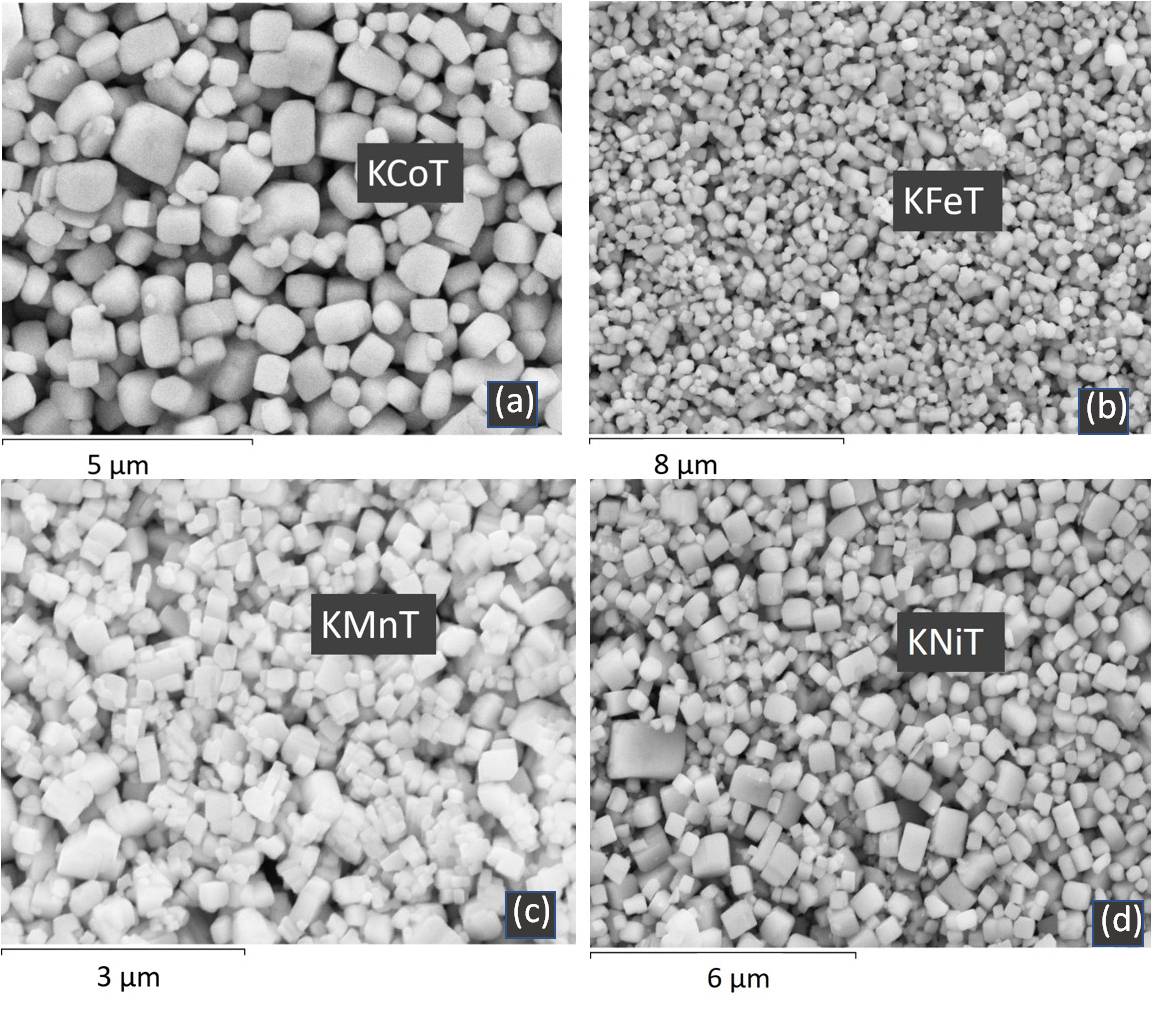}
  \caption{Scanning Electron Micrographs of the KTaO${_3}$ specimens doped with (a)Co, (b)Fe, (c)Mn and (d)Ni. The sample is highly homogeneous and shows distinctly ordered cubic structure down to few $\mu$m. All the scans were performed on  piece of pellet and showed great overlap with the expected stoichiometry of the elements.   }
  \label{Fig3}
\end{figure}
Phase purity of all the specimens were confirmed by X-Ray powder diffraction (XRD) patterns, measured using a Bruker D8 Advance diffractometer with Cu K$_{\alpha}$ source. Elemental compositions were reconfirmed by using an energy dispersive X-Ray spectrometer (Ziess Ultra Plus). Magnetization measurements were performed using a Quantum Design (MPMS-XL) SQUID magnetometer. Temperature dependent dielectric measurements were performed in the standard parallel plate geometry, using a NOVOCONTROL (Alpha-A) High Performance Frequency Analyzer. Measurements were typically done using an excitation ac signal of 1V at frequencies varying from 1 kHz to 0.5 MHz.
\begin{figure}
  \includegraphics[width=\linewidth]{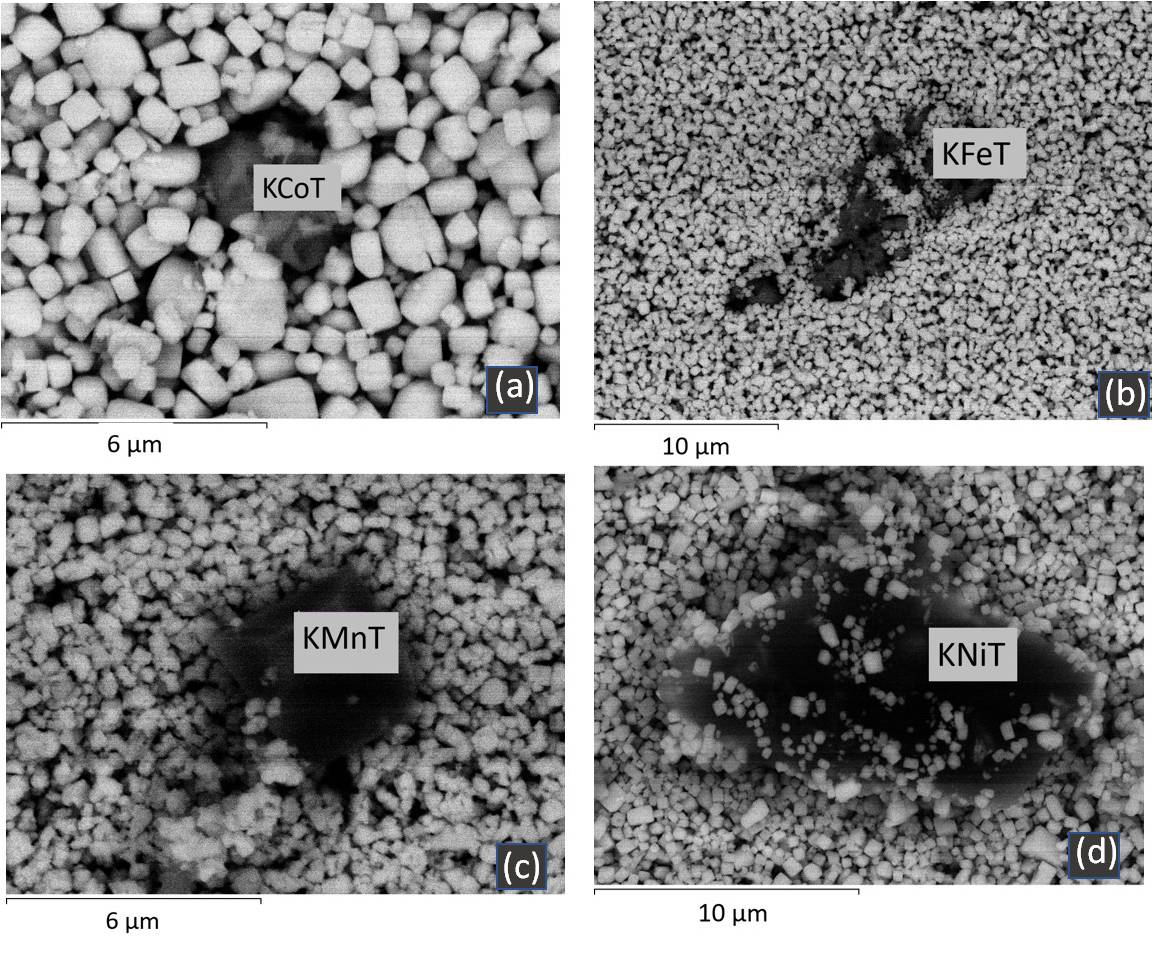}
  \caption{Backscattered images of the KTaO${_3}$ specimens doped with (a)Co, (b)Fe, (c)Mn and (d)Ni. Selected regions (in the center) which exhibited an intensity contrast corresponding to possible chemical inhomogenieties are depicted. These dark regions depicts the segregation of the transition metal oxide by exhibiting high concentration values of the respective transition metal oxide when scanned within this region, implying segregation of dopant oxides. }
  \label{Fig4}
\end{figure}
\section{ Result \& Discussions}
\begin{figure*}
  \includegraphics[width=\linewidth]{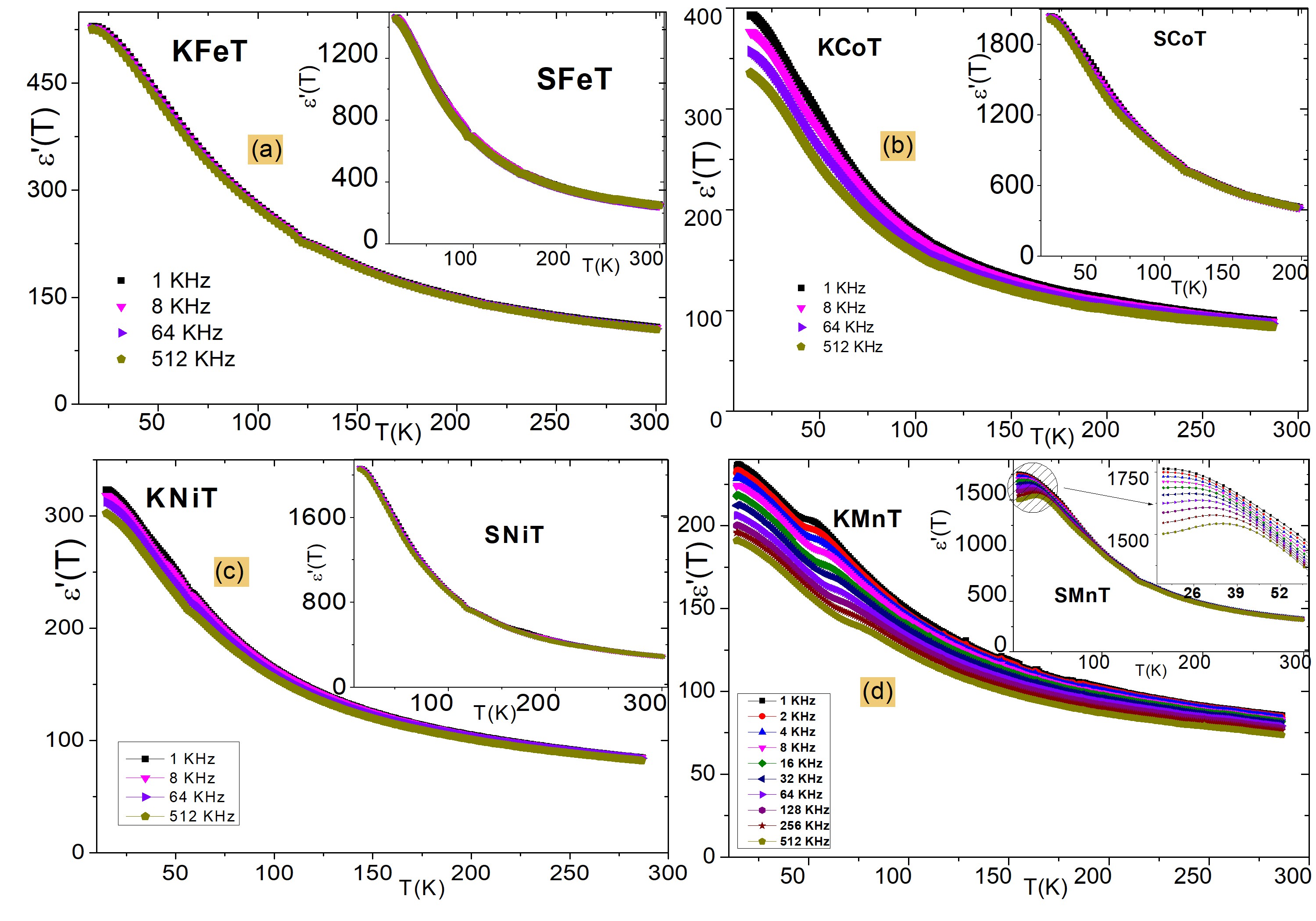}
  \caption{ Temperature dependence of the real part of the dielectric susceptibility as measured at different probing frequencies for the doped SrTiO${_3}$ (insets) and doped KTaO${_3}$ (main panels (a), (b), (c) and (d)) specimens. Only the Mn doped specimens (d) appear to exhibit an appreciable frequency dependence.}
  \label{Fig5}
\end{figure*}
\subsection{\textbf{XRD Analysis}}  The SrTiO${_3}$ and KTaO${_3}$ specimens stabilize in the cubic symmetry (space group \textit{Pm-3m}), and doping by these percentages ( 2\% and 3\% in the SMT and KMT series respectively) only results is very small changes in the lattice parameters, with no change in the crystallographic symmetry.  Here it is critical to evaluate if the XRD patterns reveals the precipitation of any of the doped transition metal oxides in the form of impurity phases.  Fig.\ref{Fig1} shows the normalized XRD patterns for all the doped SMT and KMT compounds. The 2$\theta$ range depicted here encompasses the values corresponding to the  highest intensity peaks of the transition metal oxides  (Co$_2$O$_3$, CoO, Co$_3$O$_4$,Fe$_2$O$_3$, Fe$_3$O$_4$, MnO$_2$, Mn$_2$O$_3$, Mn$_3$O$_4$, NiO or Ni$_2$O$_3$) possible as a result of Co, Fe, Mn or Ni doping. Though impurity phase detection with such small amounts of doping is typically difficult with routine XRD measurements, we have tried to glean as much information as possible by using long counting times (of the order of 6-8 hours) in these selected 2$\theta$ ranges.  In the case of the SMT series, all the doped specimens were seen to be phase pure, and no traces of any impurity oxides could be inferred from our XRD measurements. In the KMT series, the Co  [Fig.\ref{Fig1}(a)] and the Ni  [Fig.\ref{Fig1}(d)] doped specimens appear to show faint traces corresponding to the Co${_3}$O${_4}$ and NiO phases, with the Fe and Mn doped specimens  [Fig.\ref{Fig1}(b,c)] being completely free from any spurious phases.

Scanning electron Microscopy (SEM) and energy dispersive x-ray spectroscopy (EDS) measurements were done on pellets of all the specimens of the SMT and KMT series (20KV) for micro structural analysis.  Fig.\ref{Fig2} depicts the micrographs for the SMT specimens, all of which show arbitrarily shaped grains with reasonably large size distribution.  Since the dopant percentage is very low, EDS was performed at more than two dozen areas in each specimen. Though the percentage of dopants deduced from the EDS measurements are consistent with the doping values, we observed a lower percentage of Ti in some cases. This is possibly because A site substitution in SrTiO${_3}$ is not completely site selective. It has been reported, at-least for the case of Mn, that it continues to partially occupy the Ti site irrespective of the synthesis protocol used \cite{lebedev2009direct}.  For this reason, the solid solubility limit of Mn in the Sr site of SrTiO${_3}$ site has never been properly determined. Though similar studies are lacking for the other transition metals used in this study, our EDS investigations indicate that this problem appears to be generic to all the transition metals used in this investigation. 

In the case of the KMT series  [Fig.\ref{Fig3}], we observe uniform cube like microstructure, and the grains appear to have a much narrower size distribution, with clean rectangular facets and no evidence of any kind of phase clustering. The results of EDS measurements indicate no observable deviation from the actual doping percentages, suggesting that the transition metals appear to be fully incorporated in the lattice. However, considering the fact that long XRD scans suggested the possibility of spurious Co${_3}$O${_4}$ and NiO phases, and that the solubility limit of these dopants in the KTaO${_3}$ lattice is questionable, we performed additional backscattered scans on all these specimens. With the backscattering coefficient being a function of the atomic number, the contrast observed in backscattered images is known to offer a means of identifying regions of compositional inhomogeneities \cite{SEM}. After carefully sampling close to two dozen locations in the backscattering mode, we observed  regions where traces of the dopant oxides Mn$_3$O$_4$, Co$_3$O$_4$, Fe$_2$O$_3$ were seen. Identifying traces of NiO was even more difficult, and such regions could only be identified after extensive scanning. These images, where the trace impurities are evident in the form of an image contrast is depicted in Fig.\ref{Fig4}. Our exhaustive EDS measurements indicates that the solubility limit of the transition metal dopants varies from one host to the other, and also on the transition metal used. Considering the problem one is trying to address, our observations indicate that a combination of XRD, and normal and backscattering EDS measurements are imperative in evaluating the phase purity of these systems.  Our observations also indicate that all these magnetic transition metal elements have limited solubility in the SrTiO${_3}$ and KTaO${_3}$ lattices, and use of x-ray diffraction alone is not adequate to comment the phase purity of such systems. Though not included as a part of our study, high resolution synchrotron measurements could be more effective in identifying the presence of impurity phases as compared to long waiting times in laboratory x-ray diffraction. Additional information could also probably be obtained by using resonant x-ray scattering measurements.
 \subsection{\textbf{Dielectric analysis}}
\begin{figure}
	\includegraphics[width=\linewidth]{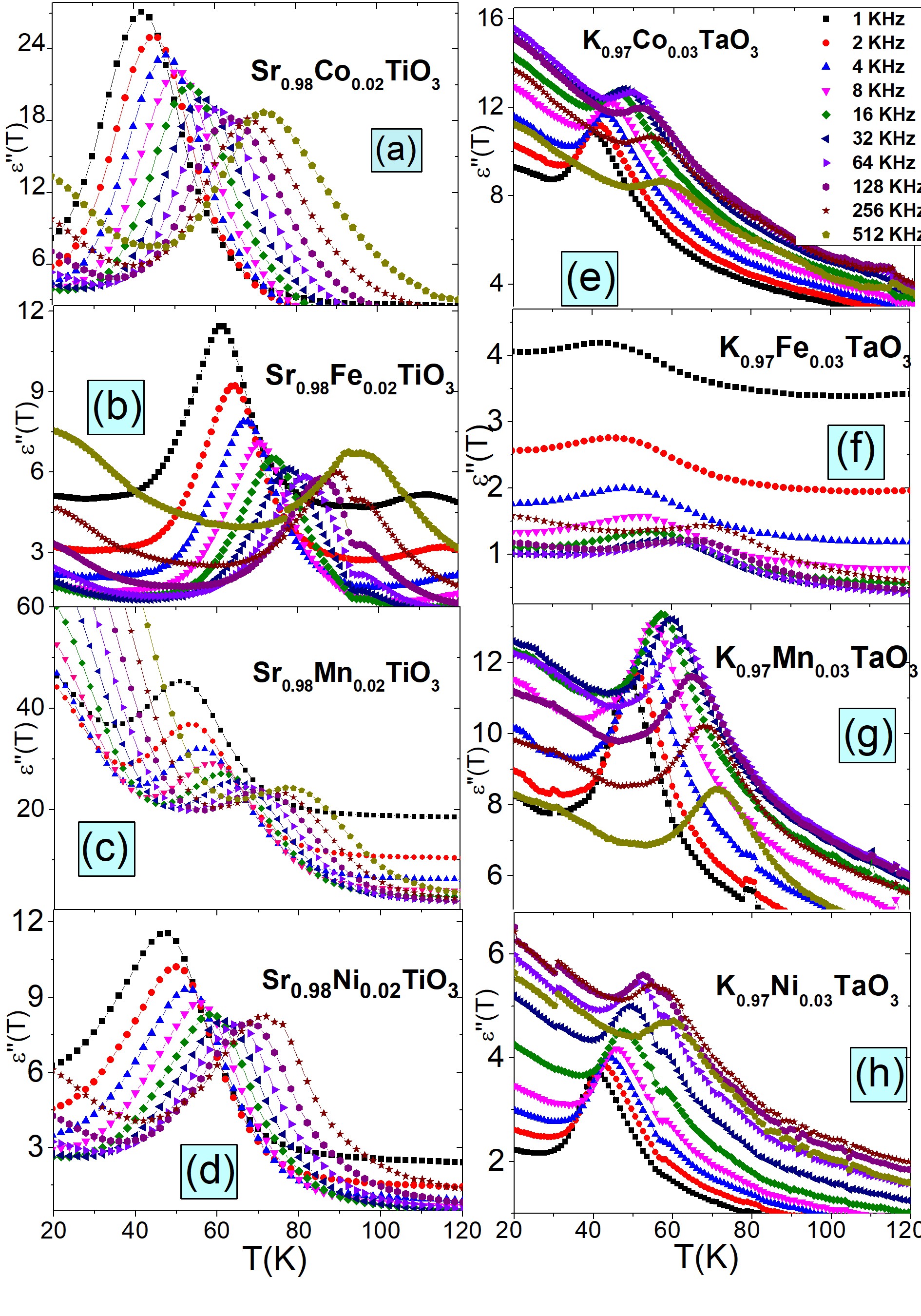}
	\caption{Temperature dependence of the imaginary part of the dielectric susceptibility as measured at different probing frequencies (1KHz to 0.5MHz) for the SrTiO${_3}$ doped with (a) Co, (b) Fe, (c) Mn and (d) Ni on the left and doped KTaO${_3}$ (e) Co, (f) Fe, (g) Mn and (h) Ni on the right.}
	\label{Fig6}
\end{figure}

SrTiO${_3}$ and KTaO${_3}$ are both known to be polarizable soft ferroelectric hosts, and the off-centered shifts of the $M{^{2+}}$ ions when doped at the 12-fold co-ordinated Sr or K positions is thought to induce electrical dipoles. These dipoles, then condense in the form of polar nano-clusters, the dynamics of which can be discerned using dielectric spectroscopy measurements. Fig.\ref{Fig5} summarizes the frequency dependent measurements of the real part of the dielectric susceptibility ($\epsilon'(T)$) over a broad temperature range for both the KMT and SMT series (inset). All the specimens exhibit a pronounced increase in $\epsilon'(T)$ as a function of decreasing temperature, as is the norm for quantum paraelectrics. We note that in pristine KTaO$_3$, a peak in the loss tangent and an anomaly in the real part of the permittivity have been reported earlier \cite{trybula,fujis}, presumably arising as a consequence of defect dipoles. The influence of the polar regions formed as a consequence of our transition metal doping is more clearly evident in the temperature dependent scans of the imaginary component of the dielectric susceptibility ($\epsilon''(T)$), as is shown in  Fig.\ref{Fig6}, where SrTiO$_{3}$ doped with (a) Co, (b) Fe, (c) Mn  and (d) Ni is depicted in the left panel, and KTaO$_{3}$ doped with (e) Co, (f) Fe, (g) Mn and (h) Ni is depicted on the right one. A broad frequency dependent peak is observed in all the specimens, with the peak in $\epsilon ''(T)$ shifting to higher temperatures with increasing frequency. This is a signature of soft ferroelectric hosts, and reflects the  dynamics of polar domains which form as a consequence of transition metal doping. The size/volume distribution of these domains is typically dependent on both the synthesis conditions as well as the nature of the dopant, as is reflected in subtle differences in both the peak position as well as its frequency dependence between the different specimens.  The dynamics associated with these polar regions are not easily seen in real part of the dielectric susceptibility $\epsilon'(T)$ due to its dramatic increase at low temperatures, associated with the renormalization of the soft mode frequency, with this increase being three orders of magnitude larger than that observed in $\epsilon''(T)$. The thermodynamic fluctuations associated with the different dopants results in an effective softening (or hardening) of this mode and dictate the effective values of $\epsilon(T)$ attained at low temperatures. 

\begin{figure}
\includegraphics[width=\linewidth]{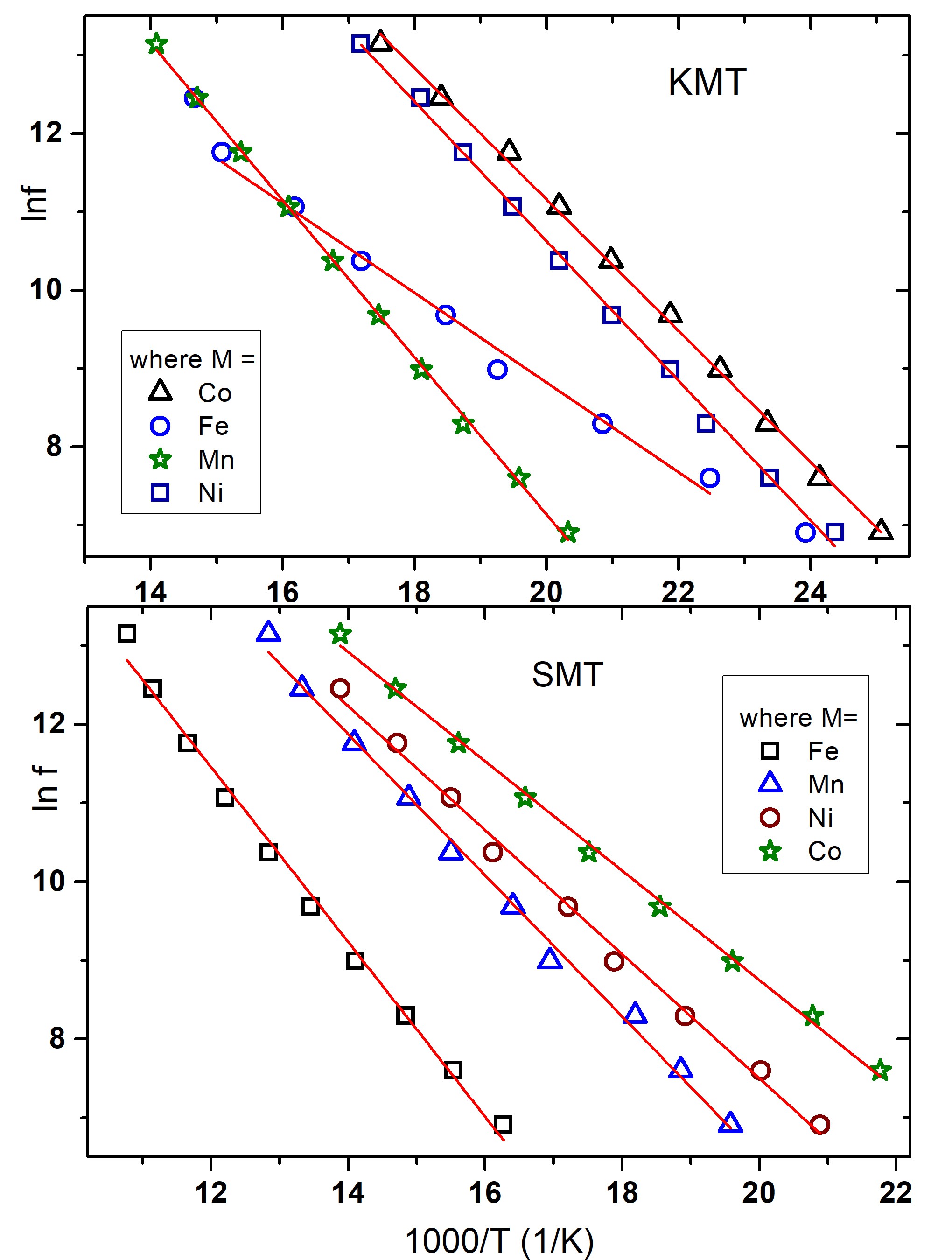}
\caption{Arrhenius fits (solid lines) to the relaxation peak observed in the doped members of the KTaO${_3}$ (top) and SrTiO${_3}$ (bottom) series. }
\label{Fig7}
\end{figure}

The frequency dependent relaxation peaks observed in the temperature dependent imaginary permittivity can be analyzed to throw light on the processes responsible for the observed relaxation. The symmetry of the perovskite lattices,in our case  SMT and KMT, allows it to hold several equivalent orientations for the dipoles formed due to disorder induced via doping. If these dipoles interact strongly and freezes into a glass-like state on lowering the temperature, one would expect to see signatures of a dynamical slowing down, which is described by a VFT formalism given as $f = f{_0} exp\frac{-E}{k{_b}[T-T{_G}]}$, where $f{_0}$, $E$, and $T{_G}$ correspond to the jump attempt frequency, the energy barrier associated with this process, and the temperature of the glass transition below which the dynamics are frozen in perspective with our measurement time scales. In contrast, if the dipoles lack co-operativity, the dipoles can hop amongst these equivalent orientations  as a consequence of thermal activation. In this case, $T{_G} \rightarrow 0$K, and the VFT equation is modified to a thermally activated Arrhenius form given by  $f = f{_0} exp\frac{-E}{k{_b}T}$. We attempted to fit the observed relaxation process to the VFT,  power law, and Arrhenius forms, and observed that best fits in all cases are given by the Arrhenius equation. This is shown in Fig.\ref{Fig7}, for all the members of the KMT (top) and SMT (bottom) series. This indicates that the density of dipoles formed by the doping of transition metal ions in the STO and KTO hosts are quite small for the doping percentages used in our investigations, and that the interaction between these dipoles are not strong enough to achieve collective freezing. There appears to be a trend in the deduced activation energies (Table I), as we see that $E_{Co}< E{_{Ni}}<E{_{Mn}}$ for both the families. Interestingly, Fe doping exhibits very different values for both the series, as the value of E$_{Fe}$ is maximum in the SMT specimens and the lowest in the KMT ones. This could presumably arise due to a substantial difference in solubility limit of Fe in both the systems.

\begin{table}
\centering
   \begin{tabular}{ | p{2cm} | p{3cm}  | p{3cm} |}
     
    \hline
 
    \textbf{Dopant} & \textbf{SMT (meV)}  &\textbf{KMT (meV)} \\ \hline
     Co & 59.35 $\pm$ 2.13  &  73.03 $\pm$ 5.40 \\ \hline
      Ni & 69.90 $\pm$ 3.11 & 77.70 $\pm$ 4.71  \\ \hline
    Mn & 79.49 $\pm$ 2.98 & 87.29 $\pm$ 4.07  \\
    \hline
    Fe & 99.95 $\pm$ 3.24 & 51.54 $\pm$ 2.84\\
    \hline
    \end{tabular}
\caption{\label{tab:table-name}The values of the activation energies deduced from the Arrhenius fit to the $\epsilon''(T)$ data }
\end {table}
    
    \begin{figure}
  \includegraphics[width=\linewidth]{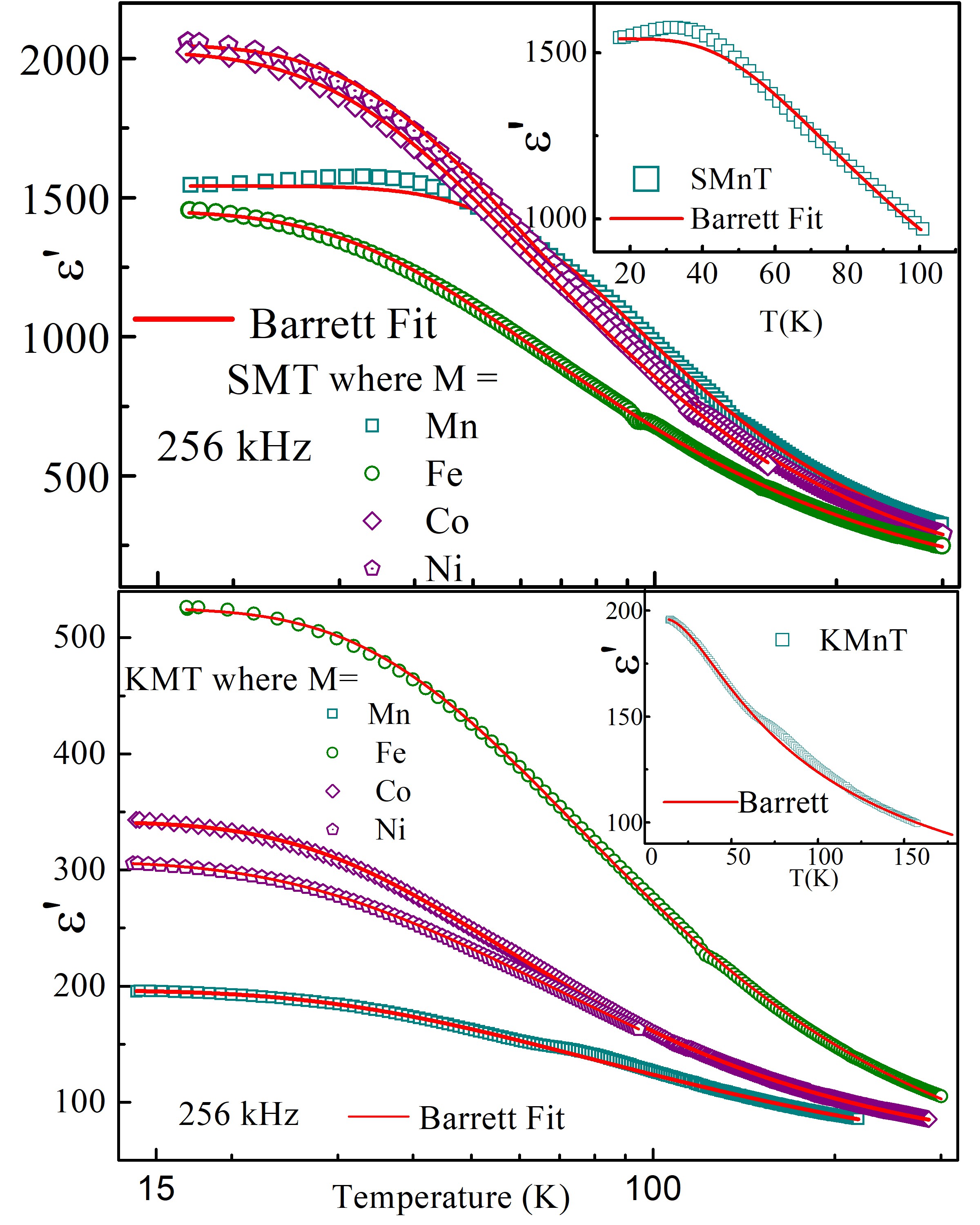}
  \caption{The measured $\epsilon$'(T) for all the doped SrTiO${_3}$ (top) and KTaO${_3}$ (bottom) specimens. The solid line corresponds to a fit using the Barrett's formalism. The insets depict deviations from the fit for the \textbf{Mn} doped specimens of both the series.}
  \label{Fig8}
\end{figure}

In doped quantum paraelectrics, the deviations from Curie-Weiss behavior is seen at low temperatures is typically ascribed to the influence of quantum fluctuations. The observed functional form has been described by the Barrett's equation  $\epsilon'$(T) = \textit{A }+ $C/[(T{_1}/2)Coth(T{_1}/2T)-T{_0}]$, where $A$ is a fitting constant (often ignored when the values of $\epsilon'(T,f)$ are sufficiently large), $C$ is the Curie constant, $T{_0}$ is equivalent to the classical Curie temperature, and $T{_1}$ represents the temperature below which quantum fluctuations overwhelm the thermal ones \cite{Barrett}. As is shown in Fig.\ref{Fig8}, good fits to the Barret's equation were obtained for the  $\epsilon'$(T) data for the Ni, Co and Fe specimens of both the SMT (top) and KMT (bottom) series, reinforcing our observation regarding the absence of a relaxor state. The fits to the dielectric susceptibility data of the Mn doped specimens of both the series as is depicted in the insets of Fig.\ref{Fig8}. This indicates that the dielectric behavior of the Mn doped specimens are distinct from those of the other transition metal doped members. This is also in broad agreement with an earlier report, where the dielectric properties of the Mn doped KTaO${_3}$ was suggested to be different from its Fe doped analogue \cite{venturini2005dipolar}. There, it was reported that whereas the Mn doped specimen depicted a peak like anomaly in $\epsilon'$(T), the Fe doped analogue did not exhibit this feature and retained a quantum paraelectric like behavior down to the lowest measured temperatures. Pressure and temperature dependent dielectric measurements were also used to suggest that the Mn doping induced dipolar entities appear to couple more strongly to the soft mode of the host quantum paraelectric lattice. 
\begin{figure}
  \includegraphics[width=\linewidth]{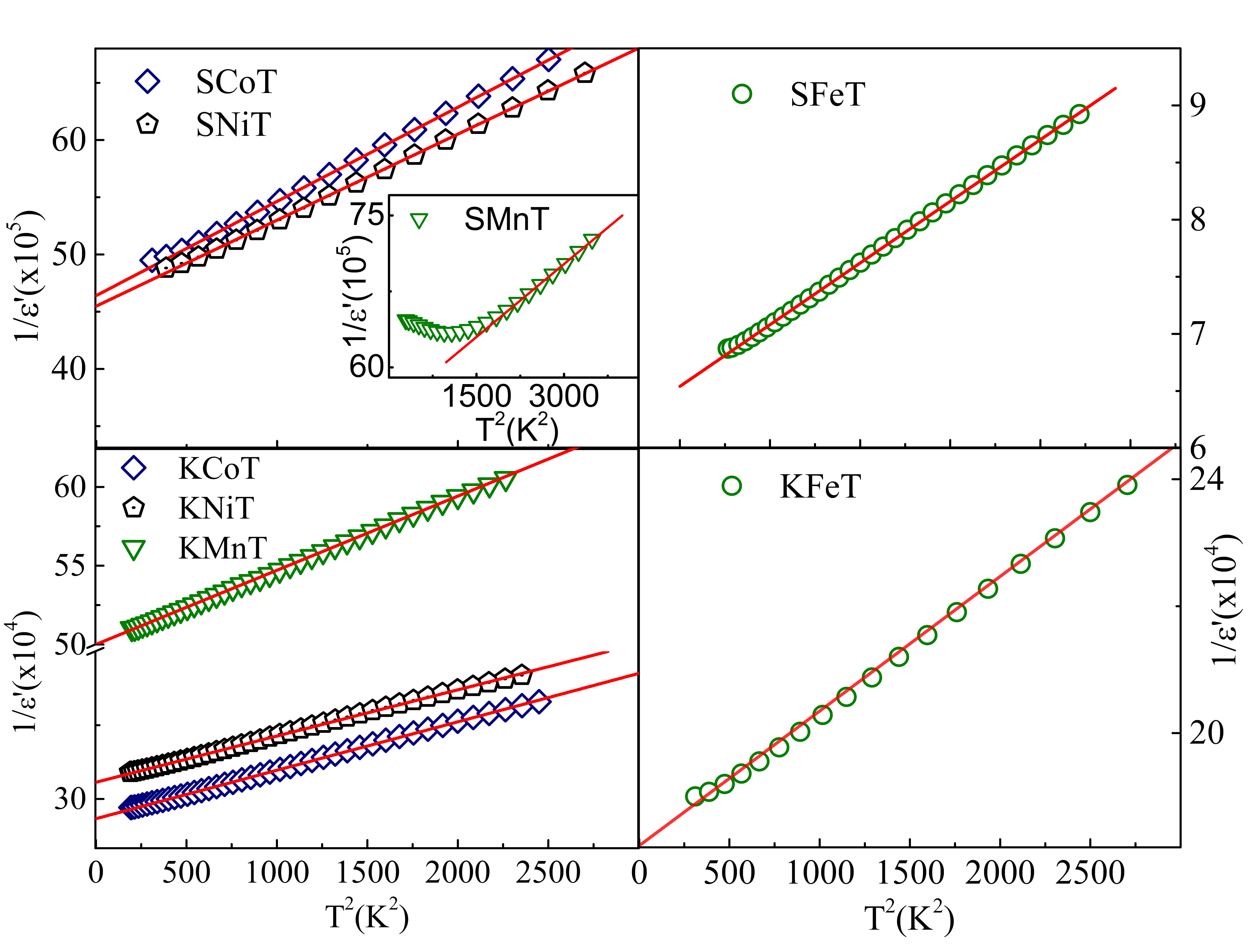}
  \caption{Inverse  permittivity as the function of the square of the temperature of doped systems. The figures (a) and (c) show the SMT and KMT systems doped with Co, Ni and Mn. The ones on the right, (b) and (d) are Fe doped SMT and KMT respectively. (The solid  straight line fitting represents the quantum critical behavior.}
      
  \label{Fig9}
\end{figure}

In the Barrett's equation, at the temperatures when T$>>$T${_1}$, the denominator term $\frac{T_{1}}{2}$  $\Big ( Coth  \frac{T_{1}}{2T} $ $\Big)$ will asymptotically approach \textit{T}, and the Barrett's equation modifies to the Curie-Weiss law at higher temperatures. On the other hand when temperature T$\rightarrow$0 K, then the dielectric permittivity varies as $\frac{C}{ \big ( \frac{T_{1}}{2} \big) - T_{o}}$ or $\frac{2C}{ \big ( T_{1} - 2T_{o} \big)}$. The positive sign of (T${_1}$-2T${_0}$) has been parametrized as a signature of the ferroelectric mode instability due to the quantum fluctuations in some earlier reports on doped quantum para-electrics \cite{tkach} \cite{wei} . The value of (T${_1}$-2T${_0}$) deduced from our fitting is positive for all the doped specimens, implying the presence of a instability of the ferroelectric mode driven by quantum fluctuations, and conclusively rules out any possibility of ferroelectricity in these doped systems and further confirms that these doped systems retain their quantum paraelectric behavior down to the lowest measured temperatures. Moreover, the magnitude of (T${_1}$-2T${_0}$) as deduced from our fits is seen to vary systematically as a function of the ionic radii of the dopant in the SMT specimens (Table II), indicating that the suppression of quantum fluctuations decreases monotonically as a function of the ionic radii. No such trend is evident in the case of the KMT specimens, presumably due to the fact that the extent of transition metals which are actually incorporated in the host lattice varies from one transition metal species to another.

In the high temperature classical regime, the inverse of the dielectric susceptibility would exhibit a Curie Weiss ( $T^{-1}$) dependence, and a defining feature of the quantum paraelectrics is the crossover to a low temperature regime where quantum fluctuations dominate. In this regime, the dielectric susceptibility is expected to vary as 1/$\epsilon'$ $\propto$ $T^d+z-2/2$, where $d$ and $z$ refer to the spatial and temporal dimensions respectively\cite{Rowley}. In the materials under consideration here, d =3, and z =1, resulting in the dielectric susceptibility varying as $T^{-2}$, as has been observed for both KTaO${_3}$ and SrTiO${_3}$ below $T\approx$50K \cite{Rowley}. Fig.\ref{Fig9} depicts the $T^2$ fit of $1/\epsilon$ for all our doped specimens below 50K, clearly indicating that doping of transition metal ions does not cause a significant change in the quantum paraelectric behavior of the host lattices\cite{prospects}, and further rules out the presence of any kind of polar order in these doped materials. We note that the variation in 1/$\epsilon$ appears to be smaller than that observed in the undoped STO and KTO specimens \cite{tkach} \cite{valant2010spin}. This could probably be due to the suppression of the quantum fluctuations as a consequence of doping. We also note that a deviation from this $T^2$ fit is observed at temperatures as high as 45K in Mn doped STO, suggesting that this dopant suppresses quantum fluctuations at a rate which is different from the other transition metals.  

The fact that the Mn doped systems behave a little differently, at least as far as its dielectric properties are concerned - could possibly due to the fact the ionic radius of Mn$^{+2}$ is larger than that of the other dopants. In the ATiO$_{3}$ perovskites, it is known that the A-O framework evokes an interstitial space, which is larger than the size of the body centred Ti$^{+4}$ ion. For instance, in BaTiO$_{3}$, a series of ferroelectric phase transitions are encountered on reducing the temperature, with each of them being associated with a different rattling mode associated with the Ti-O octahedra. On the other hand, when A is replaced by smaller cations like Ca or Sr, this rattling of Ti is hindered , and hence no ferroelectricity is observed in SrTiO$_{3}$ and CaTiO$_{3}$ down to the lowest measured temperatures\citep{kiat}\citep{ulrich}. Increasing quantum fluctuations at low temperatures also help in stabilizing the quantum paralectric state in these systems. Empirical data  suggests that smaller the size of the A site ion, larger are the quantum fluctuations, and higher is the stability of the quantum paralectric state \cite{lines}. Moreover, the critical concentration required to establish a long range or a glassy dipolar state in host quantum para-electrics also appears to be sensitive to the choice of the dopant. This has already been demonstrated earlier, where 3\% Mn doping in KTaO$_{3}$ was reported to result in a clear signature in the dielectric permittivity, whereas the same amount of Fe doping had no discernible influence \cite{venturini2005dipolar}. The fact that Mn$^{+2}$ also has the largest number of unpaired electrons among all the doped magnetic transition metal ion configurations could also play a role in influencing the complex interplay between structural considerations and quantum fluctuations in these systems.

\begin{table}

   \begin{tabular}{ | l | p{2.4cm} | p{1.2cm} | p{1.2cm} | p{1.6cm} |}
    \hline
    \textbf{KMT} & \textbf{C} &\textbf{T$_{1}$ (K)}   &\textbf{T$_{0}$ (K)} & \textbf{T$_{1}$-2T$_{0}$(K)} \\ \hline
  
     Mn & 12699$\pm$ 141 & 76.3$\pm$ 0.6 & -41.5 $\pm$ 0.9 & 159.3$\pm$1.9
       \\
    \hline
     Fe & 31557$\pm$ 86 & 107.8$\pm$ 0.7 & -6.0 $\pm$ 0.4 & 119.8$\pm$1.0
       \\
      
       \hline
   Co & 12337 $\pm$ 90 & 88.3$\pm$ 0.6 & 3.0 $\pm$ 0.5 & 94.4 $\pm$ 1.2\\   
    \hline
    Ni & 21107$\pm$ 50 & 63.7$\pm$ 0.6 & -35.6 $\pm$ 1.4 & 134.9$\pm$2.9\\
    
    \hline
    \textbf{SMT} & \textbf{C} &\textbf{T$_{1}$ (K)}   &\textbf{T$_{0}$ (K)} & \textbf{T$_{1}$-2T$_{0}$(K)} \\ \hline
    Mn & 79098$\pm$ 497& 203.4$\pm$ 1.8 & 50.9$\pm$ 1.0& 101.5$\pm$ 2.7 \\ \hline
       Fe & 71746$\pm$ 503 & 103.4$\pm$ 0.3 & 2.2 $\pm$0.1& 98.5 $\pm$ 0.4\\ \hline
    Co & 81974$\pm$ 937 & 108.9$\pm$ 1.9 & 13.7$\pm$ 1.3 & 81.5 $\pm$ 3.2\\
    \hline
    Ni & 82117$\pm$ 349  & 116.3$\pm$ 0.8 & 18.1 $\pm$ 0.5& 80.0 $\pm$ 1.3 \\
    \hline
     \end{tabular}

\caption{\label{tab:table-name}  Parameters C (Curie constant), T$_{1}$ (temperature below which quantum fluctuations overwhelm the thermal ones), T$_{0}$ (classical Curie temperature) and T$_{1}$-2T$_{0}$ evaluated from fitting of $\epsilon$'(T) using Barrett equation for KMT(top) and SMT(bottom) series. }
\end {table}

\begin{flushleft}
 \begin{table}

   \begin{tabular}{ | l | p{1.6cm} | p{1.6cm} | p{1.6cm} |}
    \hline
    \textbf{Element} & \textbf{Possible Oxides} &\textbf{Magnetic Transition Temperature}   &\textbf{Nature of Transition}  \\ \hline
    Mn & MnO$_2$ & 92K & AF \\ \hline
      & Mn$_3$O$_4$ & 41K-43K & Ferri \\ \hline
    Ni & NiO & 523K & AF \\
    \hline
     & Ni$_2$O$_3$.H$_2$O & 525K & -- \\
    \hline
    Fe &  Fe$_3$O$_4$ & 858K & Ferri \\
    \hline
      &    & 120K & Verwey \\
    \hline
      &  $\alpha$ Fe$_2$O$_3$ & 260K & Morin \\
    \hline
     &   & 950K & AF \\
    \hline
    Co& Co$_3$O$_4$ & 40K & AF \\
    \hline
      \end{tabular}
\caption{\label{tab:table-name}Magnetic transition temperatures corresponding to the different transition metal oxides which are likely to exist as possible impurities. }
\end {table}
    \end{flushleft}
    
  \subsection{\textbf{Magnetism analysis}}   
As described earlier, the presence (or the lack thereof) of a multiglass state in the doped quantum paraelectrics has been contentious due to the possibility that the magnetic spin glass like state observed in the Mn doped SrTiO${_3}$ and KTaO${_3}$ systems is extrinsic in origin. This is party due to the fact that a possible impurity oxide (Mn${_3}$O${_4}$)  has a magnetic transition in the temperature range of our interest. Moreover, it was also reported that there is no observable scaling between the onset of the magnetic anomaly and the Mn doping level \cite{valant2010spin,Kuz}. As is also observed in the structural characterization of our specimens, the solubility of magnetic ions in both SrTiO${_3}$ and KTaO${_3}$ hosts appears to be limited and kinetically hindered, making it imperative that the possibility of a magnetically frozen state is evaluated in these lattices doped with other magnetic transition metal ions.  As a reference, Table IV lists out the magnetic transition temperatures of a number of possible impurity oxides associated with the dopants being used in this study.

\begin{figure}
  \includegraphics[width=\linewidth]{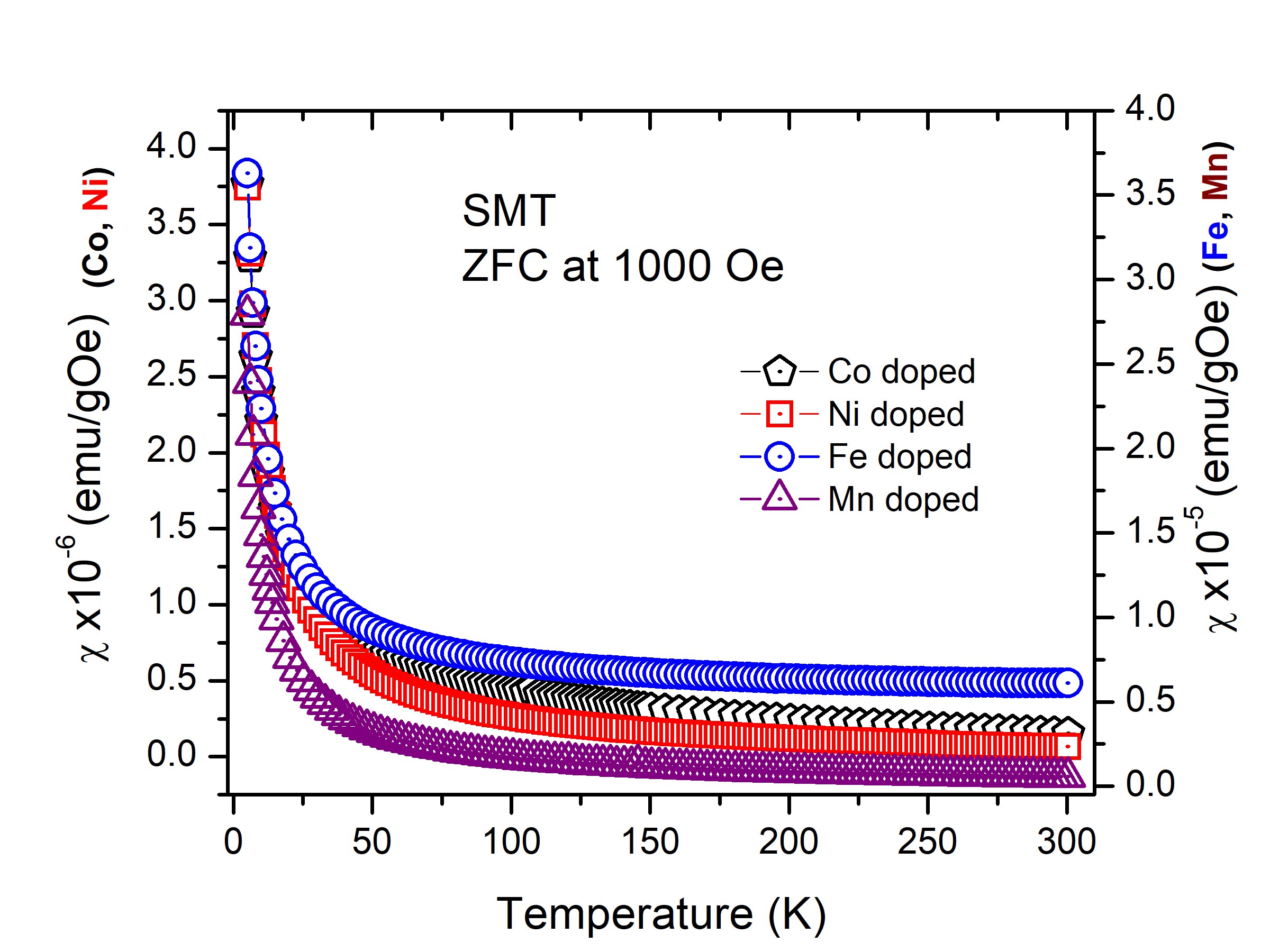}
  \caption{Temperature dependence of the dc magnetic susceptibility as measured in the doped SrTiO${_3}$ systems using the zero field cooled protocol. All the specimens appear to be paramagnetic down to the lowest measured temperatures.}
  \label{Fig10}
\end{figure}

 Fig.\ref{Fig10} depicts the magnetic measurements  performed on the SMT series in the zero field cooling mode at 1000 Oe. All the specimens (including the Mn doped one) are observed to be paramagnetic like, with no trace of any magnetic anomaly associated with long or short range magnetic order. This is in clear contradiction to the initial report in which  Mn doped SrTiO${_3}$ was suggested to be a magneto-electric multiglass.   Our results suggests that most of the Mn have been successfully incorporated into the SrTiO${_3}$ lattice, the paramagnetic contribution of which effectively overwhelms the magnetic contribution of the parasitic Mn${_3}$O${_4}$ phase. A similar scenario appears to be valid for all the doped members this series, as is evident from the fact that none of them exhibit  magnetic anomalies in the temperature range of our measurements.  
\begin{figure}
  \includegraphics[scale=0.08]{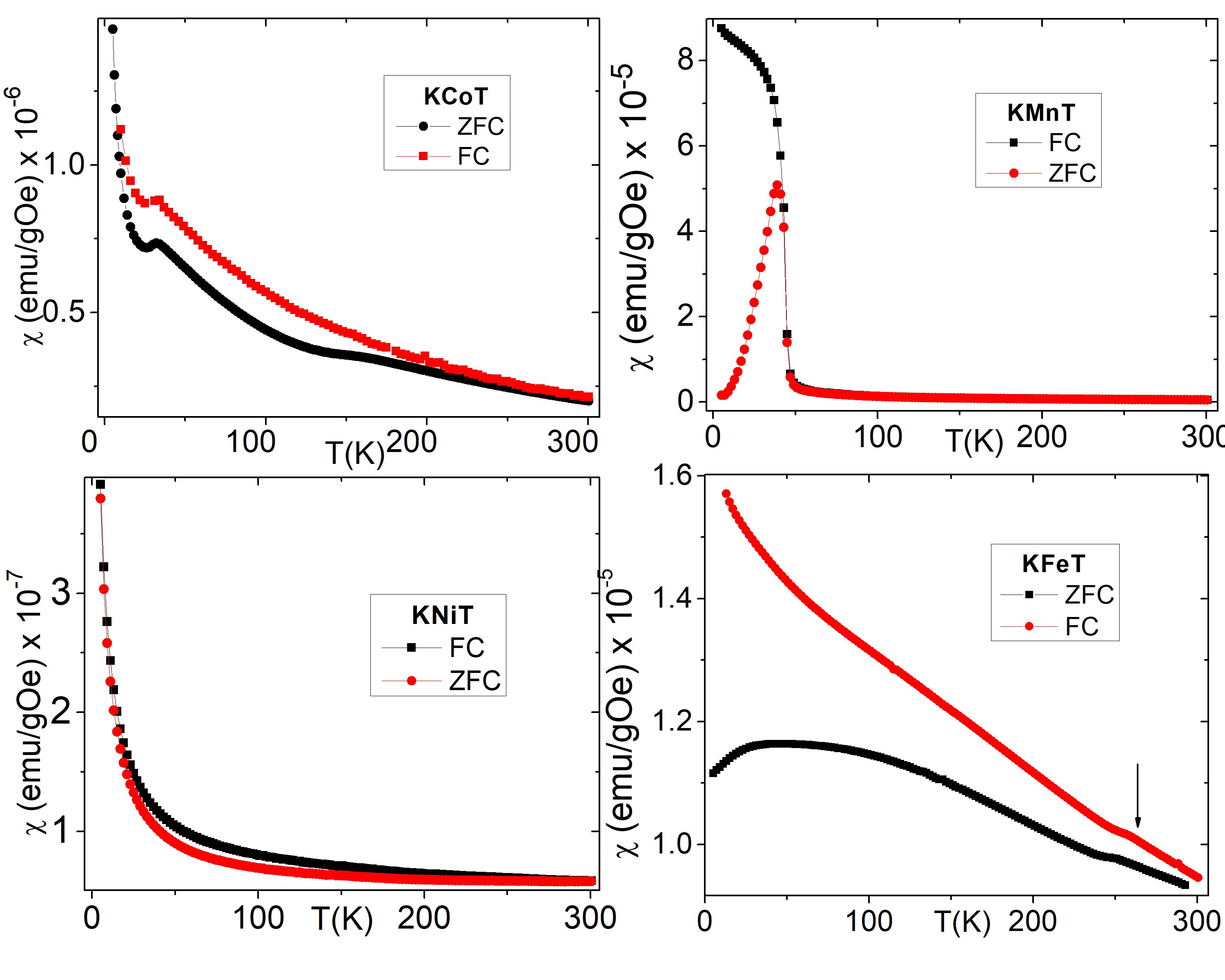}
  \caption{Temperature dependence of the dc magnetic susceptibility of the doped KTaO${_3}$ specimens as measured in the zero field cooled and field cooled protocols at 500 Oe. All these specimens exhibit signatures in the magnetization which corresponds to the magnetic transitions of one of the magnetic oxides listed in Table IV. }
  \label{Fig11}
  
\end{figure}

The magnetic measurements of the members of the KMT series as performed in the zero field cooled and field cooled protocols are shown in Fig.\ref{Fig11}, and interestingly, all the members of this series exhibit features in the magnetization. Referring to Table IV, it is evident that the Co, Mn and Fe doped specimens exhibit features corresponding to the antiferromagnetic transition of Co${_3}$O${_4}$  [Fig.\ref{Fig11}(a)], the ferrimagnetic transition of Mn${_3}$O${_4}$  [Fig.\ref{Fig11}(b)], and the Morin transition of $\alpha$-Fe${_2}$O${_3}$  [Fig.\ref{Fig11}(d)] respectively. The Ni doped specimen [Fig.\ref{Fig11}(c)] does not exhibit any feature in the magnetization, probably due to the fact that the magnetic transitions associated with the possible impurities (NiO and Ni${_2}$O${_3}$) lie above the temperature ranges of our measurements. We note that these specimens were observed to be very homogenous, and traces of impurity phases could be observed only after collecting extensive backscattering images. Though a significant amount of the magnetic dopants appear to have been incorporated into the host lattice, the doped KTaO${_3}$ systems do not appear to exhibit any additional feature in the magnetization. Thus our magnetic measurements rules out the possibility of any kind of intrinsic magnetic ordering in the doped members of both  the KMT and SMT series. Our observations clearly indicate that the magnetization of these doped specimens could vary drastically as a function of the magnetic dopants incorporated in the host lattice. The measured bulk magnetization thus reflects the competition between the intrinsic paramagnetic like susceptibility of the doped SrTiO${_3}$ and KTaO${_3}$  systems and contributions from remnant impurity oxides of the dopants. The doped quantum paraelectrics clearly remains paramagnetic down to 5K and hence any report to the contrary \cite{multiglass1,multiglass2,kleemann3,cluster} is likely to originate from extrinsic effects alone.  

\section{Conclusion}
 In summary, we have investigated the doping of magnetic transition metals (Mn, Fe, Ni, or Co) in the quantum paraelectrics SrTiO${_3}$ and KTaO${_3}$ with the aim of verifying the existence (or lack thereof) of coupled magnetic and polar glass states in these systems. Extensive structural and scanning electron microscopy measurements clearly indicate that the magnetic dopants have limited solubility in these host lattices, and traces of spurious impurities are observed even in the most well processed specimens. Our dielectric measurements show that the doping induced electrical dipoles exhibit thermally activated hopping and appear to be too weakly coupled to exhibit a critical slowing down as is expected from a frozen polar state.  Magnetic measurements indicate that all the observed magnetic signatures can be unambiguously attributed to the presence of oxides corresponding to the transition metal dopants, and that the doped  SrTiO${_3}$ and KTaO${_3}$ specimens remain paramagnetic down to the lowest measured temperatures. Our results clearly show that the doped quantum para-electrics do not harbor a multiglass state, at least in the doping concentrations investigated here, and any report to the contrary arises from extrinsic considerations alone. Though the feasibility of concomitant glassy states in the polar and magnetic sectors is an intruiging one - especially due to the possibility of exploring cross-coupled aging and rejuvenation effects - doped quantum paraelectrics are unlikely to offer a testing ground for such an exploration. 
 
\section{Acknowledgements}
The authors acknowledge Anil Prathamshetti for technical assistance in Scanning Electron Microscopy measurements. J.K. acknowledges DST India for support through PDF/2016/000911.NPDF. S.N. acknowledges DST India for support through grant no. SB/S2/CMP-048/2013 and INT/FRG/DAAD/P-249/2015.

\bibliography{Bibliography}

\end{document}